\documentclass[prl,aps,twocolumn,superscriptaddress,showpacs]{revtex4-1}

\usepackage{graphicx}% Include figure files
\usepackage{epstopdf}

\begin{document}

\title{Solitary magnons in the $S={5\over2}$ antiferromagnet CaFe$_{2}$O$_{4}$}

\author{C. Stock}
\affiliation{School of Physics and Astronomy and Centre for Science at Extreme Conditions, University of Edinburgh, Edinburgh EH9 3FD, UK}

\author{E. E. Rodriguez}
\affiliation{Department of Chemistry and Biochemistry, University of Maryland, College Park, Maryland 20742, USA}

\author{N. Lee}
\affiliation{Rutgers Center for Emergent Materials and Department of Physics and Astronomy, Rutgers University, 136 Frelinghuysen Road, Piscataway, New Jersey 08854, USA}

\author{M. A. Green}
\affiliation{School of Physical Sciences, University of Kent, Canterbury, CT2 7NH, UK}

\author{F. Demmel}
\affiliation{ISIS Facility, Rutherford Appleton Labs, Chilton, Didcot, OX11 0QX, UK}

\author{R. A. Ewings}
\affiliation{ISIS Facility, Rutherford Appleton Labs, Chilton, Didcot, OX11 0QX, UK}

\author{P. Fouquet}
\affiliation{Institute Laue-Langevin, 6 rue Jules Horowitz, Boite Postale 156, 38042 Grenoble Cedex 9, France}

\author{M. Laver}
\affiliation{School of Metallurgy and Materials, University of Birmingham, Birmingham B15 2TT, UK}

\author{Ch. Niedermayer}
\affiliation{Laboratory for Neutron Scattering, Paul Scherrer Institut, CH-5232 Villigen, Switzerland}

\author{Y. Su}
\author{K. Nemkovski}
\affiliation{J\"ulich Centre for Neuton Science JCNS, Forschungszentrum J\"ulich GmbH, Outstation at MLZ, Lichtenbergstra\ss e 1, D-85747 Garching, Germany}

\author{J. A. Rodriguez-Rivera}
\affiliation{NIST Center for Neutron Research, National Institute of Standards and Technology, 100 Bureau Drive, Gaithersburg, Maryland, 20899, USA}
\affiliation{Department of Materials Science, University of Maryland, College Park, Maryland 20742, USA}

\author{S. -W. Cheong}
\affiliation{Rutgers Center for Emergent Materials and Department of Physics and Astronomy, Rutgers University, 136 Frelinghuysen Road, Piscataway, New Jersey 08854, USA}

\date{\today}

\begin{abstract}

CaFe$_{2}$O$_{4}$ is a $S={5\over 2}$ anisotropic antiferromagnet based upon zig-zag chains having two competing magnetic structures, denoted as the A ($\uparrow \uparrow \downarrow \downarrow$) and B ($\uparrow \downarrow \uparrow \downarrow$) phases, which differ by the $c$-axis stacking of ferromagnetic stripes.  We apply neutron scattering to demonstrate that the competing A and B phase order parameters results in magnetic antiphase boundaries along $c$ which freeze on the timescale of $\sim$ 1 ns at the onset of magnetic order at 200 K.  Using high resolution neutron spectroscopy, we find quantized spin wave levels and measure 9 such excitations localized in regions $\sim$ 1-2 $c$-axis lattice constants in size.  We discuss these in the context of solitary magnons predicted to exist in anisotropic systems.  The magnetic anisotropy affords both competing A+B orders as well as localization of spin excitations in a classical magnet.  

\end{abstract}


\maketitle

Materials that display localized behavior have been studied extensively for the search of new properties including localized electronic~\cite{Anderson58:109}  and electromagnetic~\cite{John12:11} states.   In the context of magnetism, single molecular magnets have been investigated owing to the ability to tune quantum properties as well as possible device applications~\cite{Dunbar12:51,Sessoli93:365,Friedman96:76} and mesoscopic magnetic structures~\cite{Chumak15:11} have been created to confine magnetic excitations.  While localization in many of these systems is introduced through breaking up regular structures, spatial localization of energy has been known to exist in periodic structures that also contain strong nonlinear interactions.~\cite{Jartashov11:83}   Examples of such include solitary waves (solitons) or also localized breather modes.~\cite{Sievers88:61,Flach98:295,Tamga95:75}  Here we demonstrate the presence of such localized modes in a classical $S={5\over 2}$ periodic antiferromagnet where nonlinearity is introduced through magnetic anisotropy.  

\begin{figure}[t]
\includegraphics[width=8.7cm]{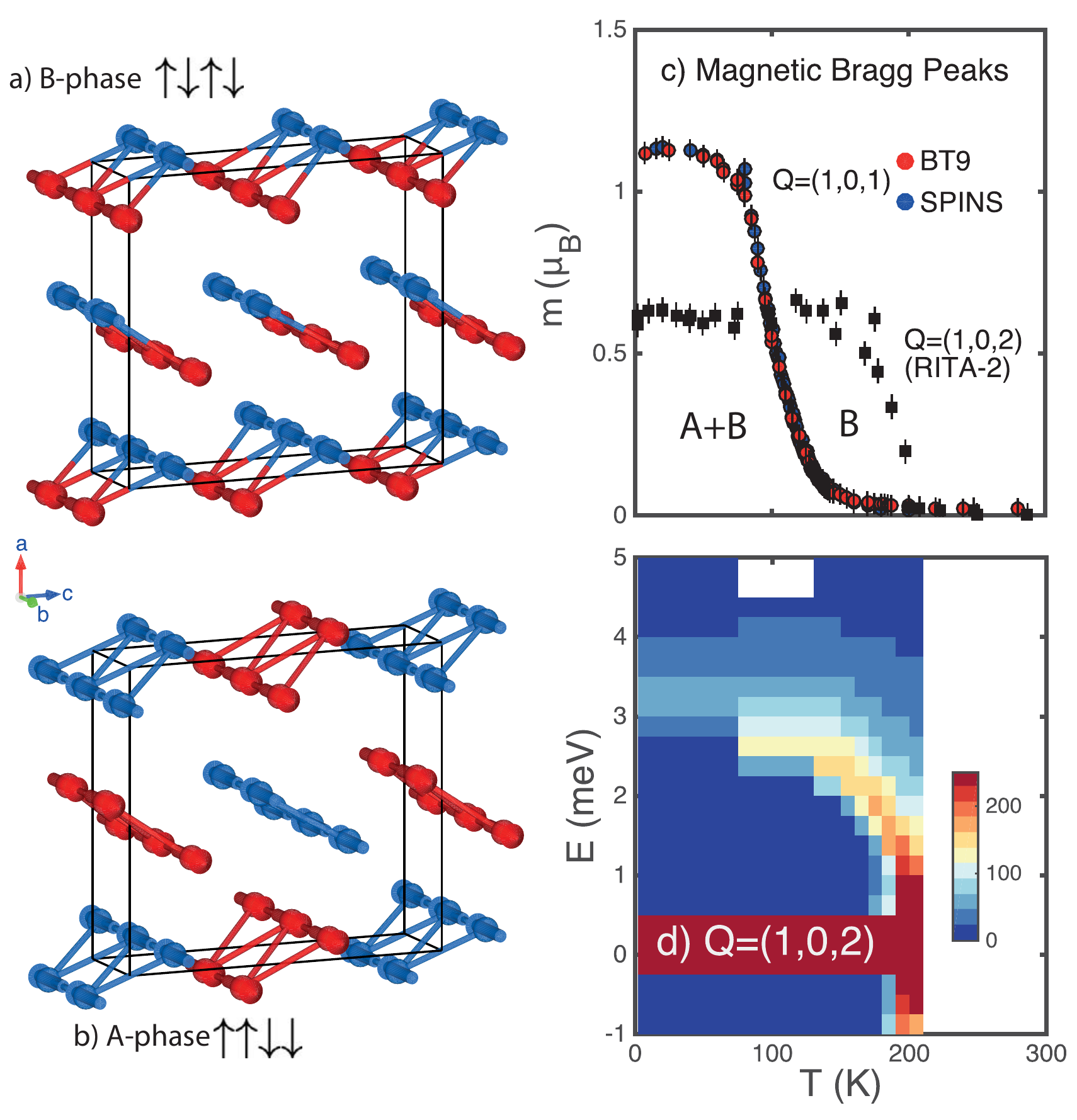}
\caption{\label{diffraction} $(a-b)$ Schematic illustration of the refined magnetic structures for the $A$ ($\uparrow \downarrow \uparrow \downarrow$) and $B$ ($\uparrow \uparrow \downarrow \downarrow$)  phases  (from powder diffraction - BT1 and polarized single crystal -DNS).  The lattice constants are $a$=9.23, $b$=3.01, and $c$=10.68 \AA.  Only the Fe atoms and their moments are shown for clarity. The magnetic moments are aligned along $b$.  $(b)$ plots the magnetic moments of the $A$ and $B$ phases as a function of temperature illustrating that both coexist at low temperatures.   $(c)$ Neutron spectroscopy data showing the opening of an anisotropy gap at 200 K which increases as temperature decreases.}
\end{figure}

CaFe$_{2}$O$_{4}$ is a $S={5\over 2}$ antiferromagnetic based upon an orthorhombic (space group 62 $Pnma$) unit cell with dimensions $a$=9.230 \AA, $b$=3.017 \AA, and $c$=10.689 \AA.~\cite{Decker57:10,Hill56:9} The magnetic structure~\cite{Allain66:9,Corliss67:160,Watanabe67:22,Bertaut10:327}  consists of two competing spin arrangements termed the $A$ and $B$ phases (illustrated in Fig. \ref{diffraction} $a-b$) which are distinguished by their $c$-axis stacking of ferromagnetic $b$-axis stripes.  The high temperature $B$ (Fig. \ref{diffraction} $a$) phase consists of stripes with antiferromagnetic alignment within the zig-zag chain network (denoted as $\uparrow \downarrow \uparrow \downarrow$).  At lower temperatures, this is replaced by the $A$ phase (Fig. \ref{diffraction} $b$) with the spins aligned parallel within the chain framework (denoted as $\uparrow \uparrow \downarrow \downarrow$).  The $A$ ($\uparrow \uparrow \downarrow \downarrow$) phase can be interpreted as the $B$ ($\uparrow \downarrow \uparrow \downarrow$) phase with an antiphase boundary along $c$. ~\cite{Corliss67:160} 

The competition between A ($\uparrow \uparrow \downarrow \downarrow$) and B ($\uparrow \downarrow \uparrow \downarrow$) phases can be motivated based on bond angles mediating superexchange interactions between the Fe$^{3+}$ spins.  The ferromagnetic alignment of the Fe$^{3+}$ spins along the $b$ axis originates from a superexchange interaction through a Fe-O-Fe bond angle of $\sim$ 86$^{\circ}$.  However, the interaction between chains within a zig-zag chain network is mediated by a bond angle of $\sim$ 100$^{\circ}$.  As outlined in Ref. \onlinecite{Shimizu03:68, Mizuno98:57} for the superexchange in the cuprates, the former bond angle is expected to be ferromagnetic while the second is marginal on the border of ferromagnetic and antiferromagnetic.  Therefore, it is not clear based on bond angles alone whether the interaction between chains, within a given zig-zag chain network, is either  ferromagnetic or antiferromagnetic and hence whether A ($\uparrow \uparrow \downarrow \downarrow$) or B ($\uparrow \downarrow \uparrow \downarrow$) is preferred.    The preferred stability of the $A$ ($\uparrow \uparrow \downarrow \downarrow$) phase has been confirmed theoretically by electronic structure calculations.~\cite{Obata13:121}  The magnetic structure in both magnetic phases is unfrustrated unlike the case in anisotropic triangular magnets.~\cite{Stock09:103}  

To investigate the phase transitions and also the dynamics of these competing magnetic order parameters, we have used neutron scattering.  The experiments were performed both on powders and single crystals of CaFe$_{2}$O$_{4}$ grown using a mirror furnace.~\cite{supp}

We first discuss the magnetic critical dynamics in CaFe$_{2}$O$_{4}$.  Diffraction results are displayed in Fig. \ref{diffraction} $c)$ illustrating the onset of $B$ (Fig. \ref{diffraction} $a$) phase ordering at 200 K, which persists to low temperatures where the $A$ (Fig. \ref{diffraction} $b$) phase smoothly onsets over a similar temperature range and then eventually dominates at low temperatures.  We have measured the temperature dependence of both $A$ and $B$ ordering using cold and thermal neutrons with both giving consistent results despite differing energy resolutions.  This indicates true temporal long range ordering in contrast to reports in systems where the magnetic ordering consists of slow relaxations.~\cite{Nambu15:115,Stock06:73,Stock10:105,Stock08:77,Murani78:41}   We have further confirmed the magnetic structure drawn in panels $(a)$ and $(b)$ on powder used to prepare the single crystal and also with polarized neutrons.  

Concomitant with the magnetic ordering, a gap opens in the magnetic excitation spectrum shown in Fig. \ref{diffraction} $(d)$ which plots a series of constant $\vec{Q}$=(1,0,2) scans taken on RITA-2 as a function of temperature.   The presence of an excitation gap is characteristic of an anisotropy term in the magnetic Hamiltonian which describes the energy needed to overcome the local alignment of the spin along the $b$ axis.  We characterize the relative size of this anisotropy in relation to the exchange energy below.

\begin{figure}[t]
\includegraphics[width=8.7cm]{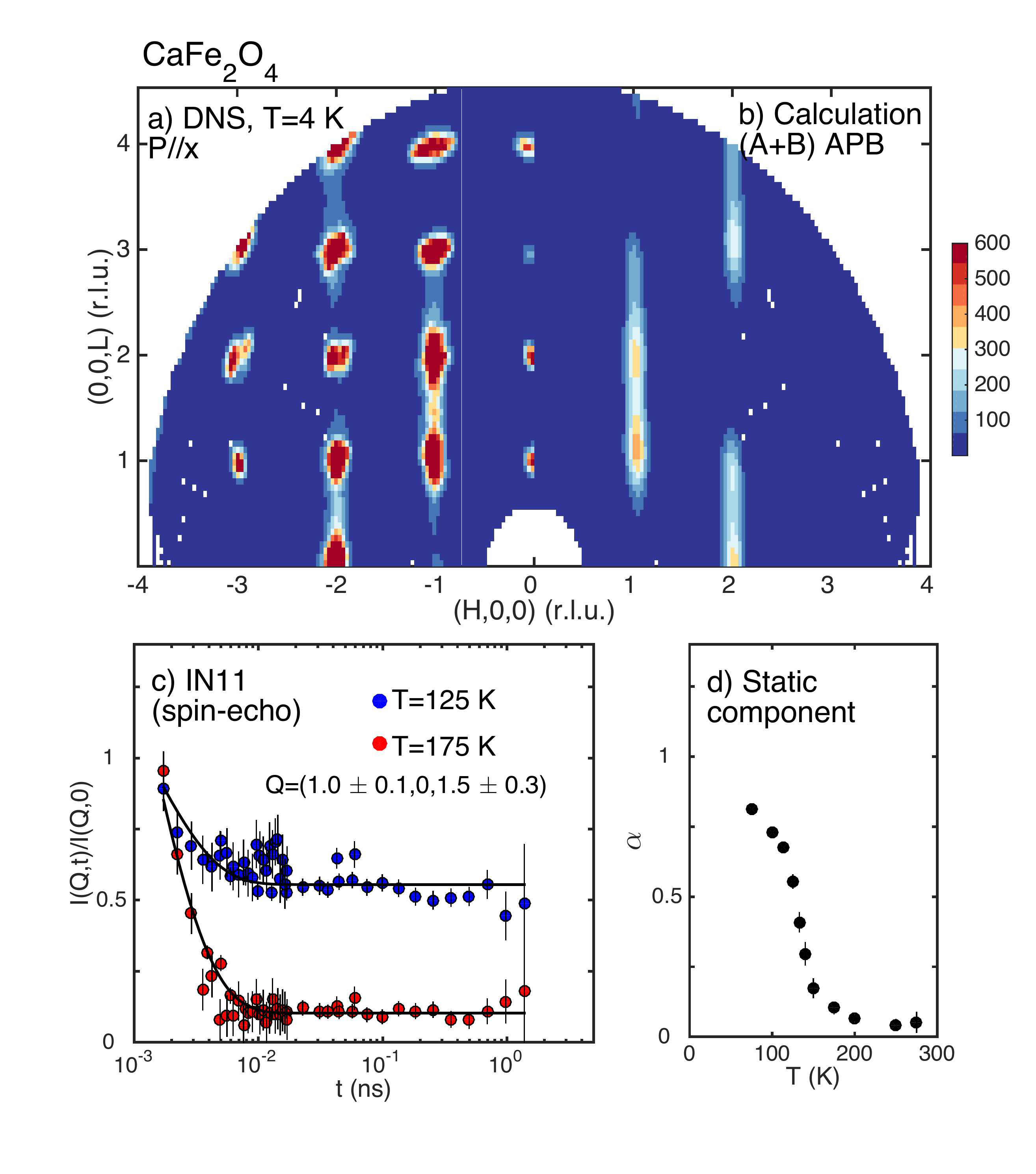}
\caption{\label{diffuse}  $(a)$ Energy integrating polarized diffuse scattering measurements confirming the magnetic nature of the diffuse rods extending along $L$.  The data is compared against a calculation in panel $(b)$ with antiphase boundaries (APB). $(c)$ shows time dependence of the real part of the normalized intermediate scattering function at 125 and 175 K.  $(d)$ plots the static component as a function of temperature.}
\end{figure}

While the magnetic ordering is long range with momentum resolution limited Bragg peaks, the total ordered moment in Fig. \ref{diffraction} $(c)$ does not reach $gS=$5 $\mu_{B}$ as expected based on Fe$^{3+}$ (S=${5\over 2}$) spins.  The missing spectral weight can be accounted for by a momentum broadened component illustrated in Fig. \ref{diffuse} $(a)$ which shows momentum broadened rods of magnetic scattering extending along $L$ from energy integrating diffraction measurements (taken using DNS at FRM2).  The polarization analysis allows us to conclude the diffuse scattering is magnetic in origin and associated with magnetic moments predominately aligned along $b$. Given that the magnetic structure also (Fig. \ref{diffraction}) consists of spins aligned along $b$ we conclude the origin comes from incomplete stacking of the $A$ and $B$ phases.   The data is compared against a calculation in Fig. \ref{diffuse} $(b)$ where we consider antiphase boundaries for both the $A$ and $B$ phases along $c$ and long-range order within the $a-b$ plane.  The cross section for this model takes the following form,

\begin{eqnarray}
I({\bf{Q}}) = \Lambda {(\gamma r_{0})^{2} \over 4} m^2 g^2 f^{2}(Q)\times \\
\left[ {\theta_{A} |F_{A}(\uparrow \downarrow \uparrow \downarrow)|^{2}+ \theta_{B}  |F_{B}(\uparrow \uparrow  \downarrow \downarrow)|^{2}}\right] \times \nonumber \\
\delta(Q_{x}-Q_{x,0})\times {\left({{\sinh(c/\xi)}\over {\cosh(c/\xi) - \cos(2 \pi L)}} \right)} \nonumber
\end{eqnarray}

\noindent where $(\gamma r_{0})^{2}$ is 0.292 b, $f(Q)$ is the Fe$^{3+}$ form factor~\cite{Brown:tables}, $g$ the Land\'e factor, $F_{A}(\uparrow \downarrow \uparrow \downarrow)|$ and $F_{B}(\uparrow \uparrow  \downarrow \downarrow)$ are the magnetic structure factors for the $A$ and $B$ phases and $\theta_{A}$ and $\theta_{B}$ are the ratios of the $A$ and $B$ phases measured in Fig. \ref{diffraction} $c)$.  $ \delta(Q_{x}-Q_{x,0})$ is the $\delta$ function along (100) indicating long-range magnetic order along $a$.  The parameter $\xi$=10 $\pm$ 2 \AA\ indicates nearest neighbor correlations along $c$.  The calibration constant $\Lambda$ was calculated based on 12 nuclear Bragg peaks from which the absolute moment in this momentum broadened part of the cross section was derived to be $m$=0.40 $\pm$ 0.10 $\mu_{B}$.   The cross section is similar to that used to analyze short-range stripe order in the cuprates and nickelates.~\cite{Tranquada96:54,Stock06:73}  Our results indicate a significant amount of the Fe$^{3+}$ spins are associated with regions which are disordered by antiphase boundaries.  

We now investigate the dynamics of these antiphase boundaries using neutron spin-echo~\cite{Mezei:nse} (Fig. \ref{diffuse} $c,d$) which probes fluctuations on the $\sim$ GHz timescale.   The data was taken on IN11 with the multidetector setup integrating in $\vec{Q}$=(1.0 $\pm$ 0.1, 0, 1.5 $\pm$ 0.3).   Spin-echo finds that the diffuse scattering consists of both a static and dynamic component between 100-200 K, with only a static component observable at temperatures below 100 K.   Example scans are shown in panel $(c)$ with solid curves fits to the relaxational form $I(Q,t)/I(Q,0)=\alpha+(1-\alpha)t^{1-\beta}\exp[-({t\over \tau})^{\beta}]$ with a temperature independent relaxation time $\tau$=1.6 $\pm$ 0.3 ps.~\cite{Pickup09:102,Pappas03:68}  The temperature independent characteristic timescale is consistent with slow relaxations investigated in random field Ising magnets with finite magnetic domains.~\cite{Villain84:52,Feng95:51,Natt88:61}  This form approaches the single relaxational lineshape when $\beta \rightarrow 1$ and we found a temperature independent $\beta$=0.90 $\pm$ 0.05 described the data well.    The stretched exponential was required to fit the data and indicates a distribution of relaxation times.  The static component $\alpha$ (measured as the baseline in Fig. \ref{diffuse} $c$) as a function of temperature is shown in Fig. \ref{diffuse} $(d)$ illustrating that the antiphase boundaries start to become static (on the timescale of $\sim$ 1 ns) at 200 K.  At 100 K, $\alpha$=0.75 indicating a substantial amount of spectral weight remains dynamic even at low temperatures. 

\begin{figure}[t]
\includegraphics[width=8.7cm]{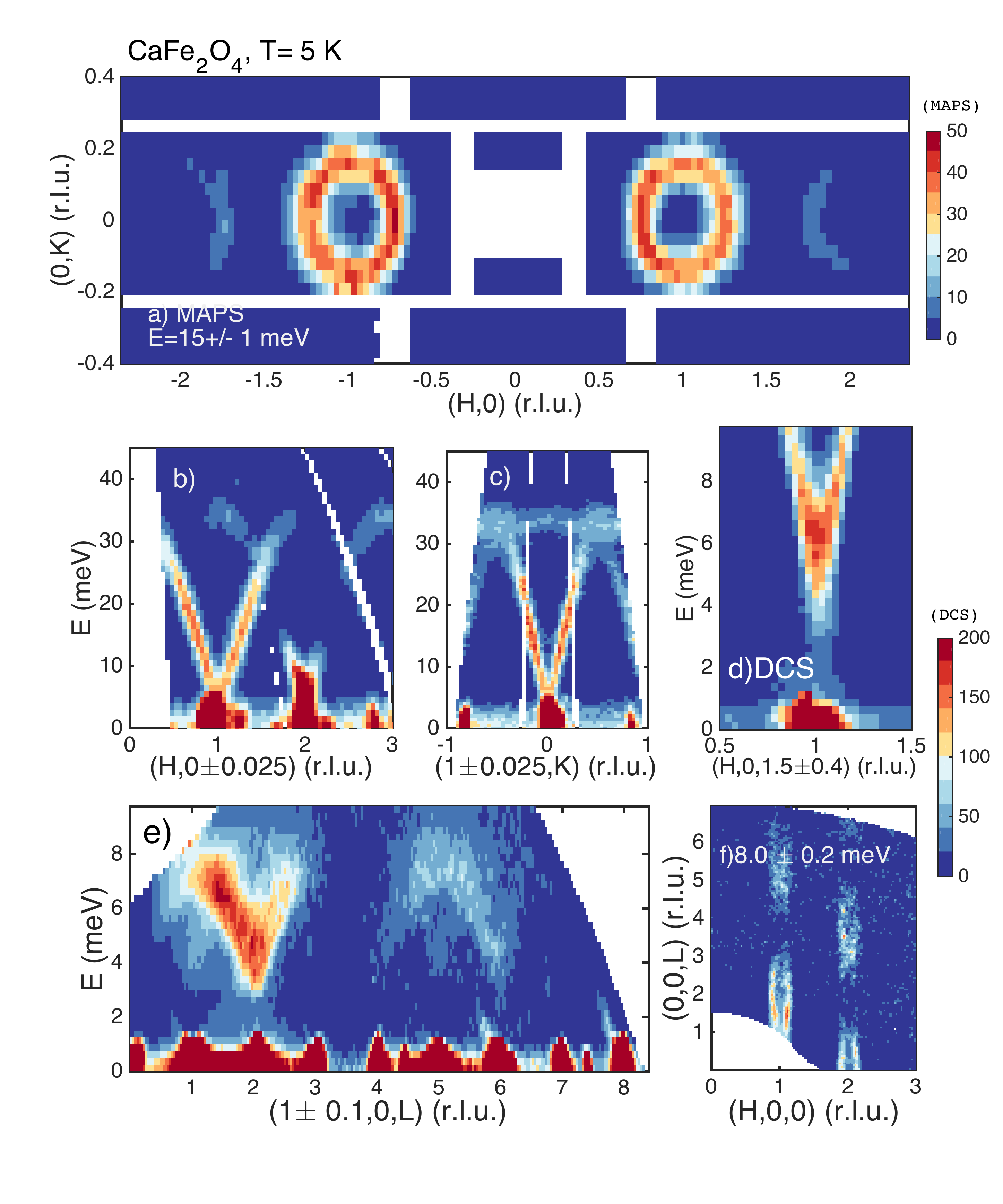}
\caption{\label{inelastic}  $(a)$ Constant energy slice illustrating rings of magnetic scattering in the (H,K) plane taken on MAPS.  $(b,c)$ display strong dispersion along the H and K directions from MAPS.  $(d)$ illustrates high resolution DCS data showing a low temperature anisotropy gap. $e)$ shows the dispersion of the excitations along $c$ with a low energy constant energy slice illustrated in $(f)$ in the (H,0,L) plane.}
\end{figure}

To understand the microscopic origin for the antiphase boundaries, we investigate the spatial spin coupling along the three crystallographic directions by measuring the magnetic dispersion curves (Fig. \ref{inelastic} at T= 5 K).  Figure \ref{inelastic} $(a)$ shows a constant energy (E=15 $\pm$ 1 meV) slice taken on MAPS illustrating well defined rings of scattering in the (H,K) plane.  Dispersion curves along the $a$ and $b$ crystallographic axes are plotted in panels $(b)$ and $(c)$ illustrating strong and nearly isotropic dispersion in this plane indicative of strong exchange coupled spins within the $a-b$ plane.  Higher resolution data (panel $d$) shows the presence of an anisotropy gap of $\Delta$=5.0  $\pm$ 0.3 meV.  Fits to dispersion relations along $a$ and $b$ find a spin-wave velocity of $\hbar v$ = 78 $\pm$ 5 meV \AA.  The results along the (0,0,L) direction are different and shown in panels $(e)$ where a significantly smaller dispersion is found with a fitted $J_{c}/J_{ab}$=0.14 $\pm$ 0.03.  This is confirmed by a constant energy slice (8.0 $\pm$ 2.0 meV) $(f)$ which displays rods of scattering following the structure factor used to model the $A$ and $B$ diffuse scattering above.  Based on these dispersion curves, we find the spin coupling in CaFe$_{2}$O$_{4}$ is strongly two dimensional with weak spin coupling along the (0,0,L) direction.    The strong two dimensionality characterized by the large difference in spin exchange ($J_{c}/J_{ab}$) facilitates the formation of antiphase boundaries along $c$ discussed above.  

\begin{figure}[t]
\includegraphics[width=9.2cm]{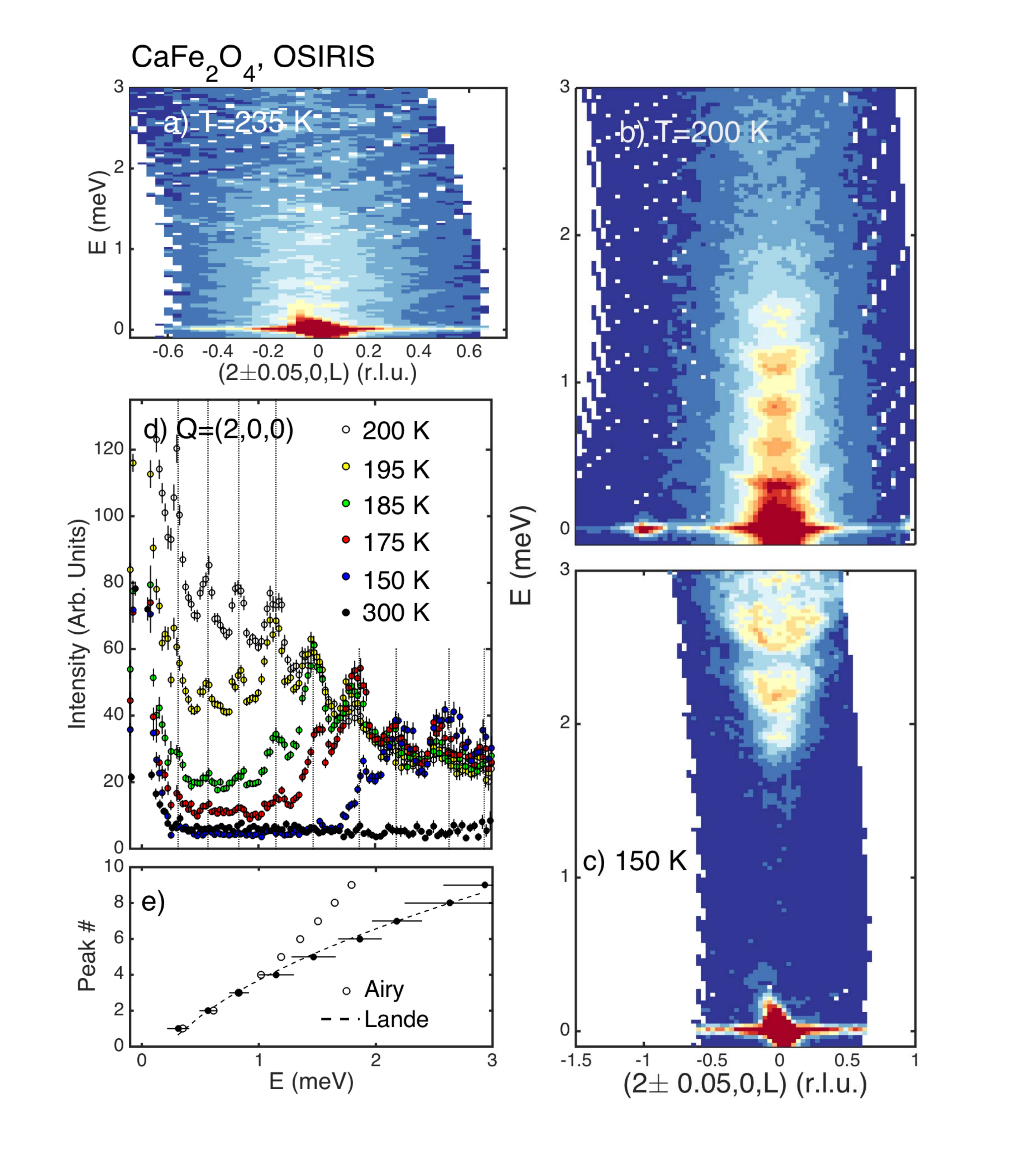}
\caption{\label{osiris}  High resolution spectroscopy illustrating the presence of countable magnetic excitations. $(a)$ shows temporally and momentum broadened excitations at high temperatures.  $(b-c)$ illustrate quantized spin excitations at 200 and 150 K with a summary of a series of temperatures plotted in $(d)$.  The peak position is plotted against energy in $(e)$ and compared against expectations based on the Airy analysis and a heuristic model based on the Land\'e interval theorem and also proposed by theory for anisotropic classical chains.}
\end{figure}

We now investigate the spin excitations using the fine energy resolution afforded by neutron backscattering.  High resolution temperature dependent spectroscopy of the magnetic excitations is illustrated in Fig. \ref{osiris} with an elastic resolution of $\delta E$=0.025 meV (full-width at half maximum).  At high temperatures of 235 K (panel $a$) momentum and temporally broadened correlations are observed consistent with paramagnetic critical scattering for the formation of the $B$ phase.  At high temperatures, thermal fluctuations dominate the dynamics.

However, at 200 K (panel $b$), where static antiphase boundaries begin to form (Fig. \ref{diffuse} $d$),  this is replaced by a series of discrete quantized excitations.  The heirarchy of excitations persists for temperatures below the magnetic ordering temperature of 200 K and is shown in panel $(c)$ at 150 K.   A summary of a series of temperatures from 150 K-300 K is shown in panel $d)$ where 9 discrete energies are observed.   The spectral weight distribution as a function of temperature tracks the recovery of the anisotropy gap studied with coarser energy resolution in Fig. \ref{diffraction}.   The discrete nature is most clearly resolved in scans along $L$ given the weak spin coupling and hence softer spin wave dispersion along this direction.  

We now discuss the origin of the quantized spin waves observed below T$_{N}$ shown in Fig. \ref{osiris}.  A possible mechanism for quantized magnetic fluctuations is through a binding interaction as postulated to exist in CoNb$_{2}$O$_{6}$~\cite{Coldea10:327,Morris14:112} where discrete excitations are stabilized by interactions between spinons.  Our data is inconsistent with a theory describing this situation based upon a Schr\"odinger equation where the potential energy is proportional to the separation of the spinons (in the case of S=${1\over 2}$).  Confirming this conclusion, Fig. \ref{osiris} $(e)$ shows a plot of the energy position of each mode compared against the negative roots of the Airy function which are the solutions of this Schr\"odinger equation.  The data diverges markedly from the prediction at higher energy and larger peak number ($N$) indicating the discrete nature does not originate from interactions.

Another possibility is that the discrete excitations originate from a localized spin object  where the underlying Hamiltonian is $H=J\sum_{n} \vec{S}_{n} \cdot \vec{S}_{n+1}$, with the sum over a finite number of spins.  Such a Hamiltonian, which is bilinear in angular momentum, is subject to the Land\'e interval rule which states that the interval between two neighboring levels is proportional to the higher $j$ (quantum number associated with the total angular momentum operator) value of the pair.~\cite{Judd:book}  Hence, for increasing energy and hence $j$, assuming antiferromagnetic interactions, the level spacing is predicted to increase in agreement with the data.  In panel $(e)$, we fit the peak position to the heuristic form $N=\Pi \sqrt{E}+ \Phi$, where $\Pi$ and $\Phi$ are fitting constants, $E$ is the energy position, and $N$ is the peak integer.  This heuristic model motivated by the interval theorem provides a good description of the data over the 9 observed peaks implicating localized regions of confined classical spins as the origin of the discrete spin waves.   

The discrete excitations are only observed below T$_{N}$=200 K where static magnetic order and antiphase boundaries are present.  An estimate of the size of these localized regions is given by the correlation length derived from the diffraction data presented above in Fig. \ref{diffuse} with a lengthscale of only 1-2 unit cells along $c$.   The size of these localized regions in the $a-b$ plane is determined by the resolution ($\sim$ 100 \AA) indicating that they are highly anisotropic spatially.  The observed modes are analogous to discrete breathers as they are spatially localized, time periodic, and stable excitations.   However, unlike localized nonlinear solitons which can propagate without dampening, breathers are known to decay~\cite{Sievers88:61,Sato04:432,Sakaguchi05:72}, consistent with the temperature independent relaxation on the 1 GHz timescale measured in Fig. \ref{diffuse}.   

Stabilizing localized excitations has been predicted in lattices which have nonlinear terms in the energy expansion.  In the context of the antiferromagnetic Hamiltonian, this can occur through anisotropic energies originating from a distorted octahedra around the magnetic Fe$^{3+}$ ion.  An analogous case has been considered in a number of studies and most notably in a classic spin-chain~\cite{Flach08:497,Mikeska78:11,Haldane83:50,Fogedby80:13,Haldane82:15}.  Theoretical studies~\cite{Pushkarov77:81,Pushkarov78:85,Pushkarov79:93,Mikeska91:40} found the semiclassical (high spin) anisotropic Hamiltonian to map onto the discrete nonlinear Schr\"odinger equation and to support soliton excitations where the energy scaled as $\propto N^{2}$ - the result argued above based on the Land\'e interval rule.   A similar increase in frequency spacing has also been modelled in nonlinear electronic circuits~\cite{Marquie95:51}, lattices~\cite{Denardo92:68}, and waveguides~\cite{Eisenberg98:81} which also are based upon similar nonlinear equations. 

Localized antiferromagnetic excitations have been observed through driving large amplitude spin-waves in low-dimensional antiferromagnets.~\cite{Sato04:432}  Countable excitations were measured as a step in the time dependence of the emission signal.~\cite{Sato05:71}  The anisotropy in CaFe$_{2}$O$_{4}$ allows such localized countable excitations to exist even under mild perturbations with neutrons.

In summary, we have reported the magnetic excitations and transitions in CaFe$_{2}$O$_{4}$.   We find  competing anisotropic order parameters which support countable spin excitations.  CaFe$_{2}$O$_{4}$ therefore displays solitary excitations confining energy locally.

\begin{acknowledgements}
This work was supported by the EPSRC, Carnegie Trust for the Universities of Scotland, Royal Society of London, Royal Society of Edinburgh, EU-NMi3, NSF (DMR-1508249), and the Swiss spallation neutron source (SINQ) (Paul Scherrer Institute, Villigen, Switzerland). The work at Rutgers University was supported by the DOE under Grant No. DOE: DE-FG02-07ER46382..  Further details about open access data is given the supplementary information.
\end{acknowledgements}

%\thebibliography{}

%\bibliography{CFO_bib}

\begin{thebibliography}{52}%
\makeatletter
\providecommand \@ifxundefined [1]{%
 \@ifx{#1\undefined}
}%
\providecommand \@ifnum [1]{%
 \ifnum #1\expandafter \@firstoftwo
 \else \expandafter \@secondoftwo
 \fi
}%
\providecommand \@ifx [1]{%
 \ifx #1\expandafter \@firstoftwo
 \else \expandafter \@secondoftwo
 \fi
}%
\providecommand \natexlab [1]{#1}%
\providecommand \enquote  [1]{``#1''}%
\providecommand \bibnamefont  [1]{#1}%
\providecommand \bibfnamefont [1]{#1}%
\providecommand \citenamefont [1]{#1}%
\providecommand \href@noop [0]{\@secondoftwo}%
\providecommand \href [0]{\begingroup \@sanitize@url \@href}%
\providecommand \@href[1]{\@@startlink{#1}\@@href}%
\providecommand \@@href[1]{\endgroup#1\@@endlink}%
\providecommand \@sanitize@url [0]{\catcode `\\12\catcode `\$12\catcode
  `\&12\catcode `\#12\catcode `\^12\catcode `\_12\catcode `\%12\relax}%
\providecommand \@@startlink[1]{}%
\providecommand \@@endlink[0]{}%
\providecommand \url  [0]{\begingroup\@sanitize@url \@url }%
\providecommand \@url [1]{\endgroup\@href {#1}{\urlprefix }}%
\providecommand \urlprefix  [0]{URL }%
\providecommand \Eprint [0]{\href }%
\providecommand \doibase [0]{http://dx.doi.org/}%
\providecommand \selectlanguage [0]{\@gobble}%
\providecommand \bibinfo  [0]{\@secondoftwo}%
\providecommand \bibfield  [0]{\@secondoftwo}%
\providecommand \translation [1]{[#1]}%
\providecommand \BibitemOpen [0]{}%
\providecommand \bibitemStop [0]{}%
\providecommand \bibitemNoStop [0]{.\EOS\space}%
\providecommand \EOS [0]{\spacefactor3000\relax}%
\providecommand \BibitemShut  [1]{\csname bibitem#1\endcsname}%
\let\auto@bib@innerbib\@empty
%</preamble>
\bibitem [{\citenamefont {Anderson}(1958)}]{Anderson58:109}%
  \BibitemOpen
  \bibfield  {author} {\bibinfo {author} {\bibfnamefont {P.~W.}\ \bibnamefont
  {Anderson}},\ }\href@noop {} {\bibfield  {journal} {\bibinfo  {journal}
  {Phys. Rev.}\ }\textbf {\bibinfo {volume} {109}},\ \bibinfo {pages} {1492}
  (\bibinfo {year} {1958})}\BibitemShut {NoStop}%
\bibitem [{\citenamefont {John}(2012)}]{John12:11}%
  \BibitemOpen
  \bibfield  {author} {\bibinfo {author} {\bibfnamefont {S.}~\bibnamefont
  {John}},\ }\href@noop {} {\bibfield  {journal} {\bibinfo  {journal} {Nature
  Materials}\ }\textbf {\bibinfo {volume} {11}},\ \bibinfo {pages} {997}
  (\bibinfo {year} {2012})}\BibitemShut {NoStop}%
\bibitem [{\citenamefont {Dunbar}(2012)}]{Dunbar12:51}%
  \BibitemOpen
  \bibfield  {author} {\bibinfo {author} {\bibfnamefont {K.~R.}\ \bibnamefont
  {Dunbar}},\ }\href@noop {} {\bibfield  {journal} {\bibinfo  {journal} {Inorg.
  Chem.}\ }\textbf {\bibinfo {volume} {51}},\ \bibinfo {pages} {12055}
  (\bibinfo {year} {2012})}\BibitemShut {NoStop}%
\bibitem [{\citenamefont {Sessoli}\ \emph {et~al.}(1993)\citenamefont
  {Sessoli}, \citenamefont {Gatteschi}, \citenamefont {Caneschi},\ and\
  \citenamefont {Novak}}]{Sessoli93:365}%
  \BibitemOpen
  \bibfield  {author} {\bibinfo {author} {\bibfnamefont {R.}~\bibnamefont
  {Sessoli}}, \bibinfo {author} {\bibfnamefont {D.}~\bibnamefont {Gatteschi}},
  \bibinfo {author} {\bibfnamefont {A.}~\bibnamefont {Caneschi}}, \ and\
  \bibinfo {author} {\bibfnamefont {M.~A.}\ \bibnamefont {Novak}},\ }\href@noop
  {} {\bibfield  {journal} {\bibinfo  {journal} {Nature}\ }\textbf {\bibinfo
  {volume} {365}},\ \bibinfo {pages} {141} (\bibinfo {year}
  {1993})}\BibitemShut {NoStop}%
\bibitem [{\citenamefont {Friedman}\ \emph {et~al.}(1996)\citenamefont
  {Friedman}, \citenamefont {Sarachik}, \citenamefont {Tejada},\ and\
  \citenamefont {Ziolo}}]{Friedman96:76}%
  \BibitemOpen
  \bibfield  {author} {\bibinfo {author} {\bibfnamefont {J.~R.}\ \bibnamefont
  {Friedman}}, \bibinfo {author} {\bibfnamefont {M.~P.}\ \bibnamefont
  {Sarachik}}, \bibinfo {author} {\bibfnamefont {J.}~\bibnamefont {Tejada}}, \
  and\ \bibinfo {author} {\bibfnamefont {R.}~\bibnamefont {Ziolo}},\
  }\href@noop {} {\bibfield  {journal} {\bibinfo  {journal} {Phys. Rev. Lett.}\
  }\textbf {\bibinfo {volume} {76}},\ \bibinfo {pages} {3830} (\bibinfo {year}
  {1996})}\BibitemShut {NoStop}%
\bibitem [{\citenamefont {Chumak}\ \emph {et~al.}(2015)\citenamefont {Chumak},
  \citenamefont {Vasyuchka}, \citenamefont {Serga},\ and\ \citenamefont
  {Hillebrands}}]{Chumak15:11}%
  \BibitemOpen
  \bibfield  {author} {\bibinfo {author} {\bibfnamefont {A.~V.}\ \bibnamefont
  {Chumak}}, \bibinfo {author} {\bibfnamefont {V.~I.}\ \bibnamefont
  {Vasyuchka}}, \bibinfo {author} {\bibfnamefont {A.~A.}\ \bibnamefont
  {Serga}}, \ and\ \bibinfo {author} {\bibfnamefont {B.}~\bibnamefont
  {Hillebrands}},\ }\href@noop {} {\bibfield  {journal} {\bibinfo  {journal}
  {Nature Physics}\ }\textbf {\bibinfo {volume} {11}},\ \bibinfo {pages} {453}
  (\bibinfo {year} {2015})}\BibitemShut {NoStop}%
\bibitem [{\citenamefont {Kartashov}\ \emph {et~al.}(2011)\citenamefont
  {Kartashov}, \citenamefont {Malomed},\ and\ \citenamefont
  {Torner}}]{Jartashov11:83}%
  \BibitemOpen
  \bibfield  {author} {\bibinfo {author} {\bibfnamefont {Y.~V.}\ \bibnamefont
  {Kartashov}}, \bibinfo {author} {\bibfnamefont {B.~A.}\ \bibnamefont
  {Malomed}}, \ and\ \bibinfo {author} {\bibfnamefont {L.}~\bibnamefont
  {Torner}},\ }\href@noop {} {\bibfield  {journal} {\bibinfo  {journal} {Rev.
  Mod. Phys.}\ }\textbf {\bibinfo {volume} {83}},\ \bibinfo {pages} {247}
  (\bibinfo {year} {2011})}\BibitemShut {NoStop}%
\bibitem [{\citenamefont {Sievers}\ and\ \citenamefont
  {Takeno}(1988)}]{Sievers88:61}%
  \BibitemOpen
  \bibfield  {author} {\bibinfo {author} {\bibfnamefont {A.~J.}\ \bibnamefont
  {Sievers}}\ and\ \bibinfo {author} {\bibfnamefont {S.}~\bibnamefont
  {Takeno}},\ }\href@noop {} {\bibfield  {journal} {\bibinfo  {journal} {Phys.
  Rev. Lett.}\ }\textbf {\bibinfo {volume} {61}},\ \bibinfo {pages} {970}
  (\bibinfo {year} {1988})}\BibitemShut {NoStop}%
\bibitem [{\citenamefont {Flach}\ and\ \citenamefont
  {Willis}(1998)}]{Flach98:295}%
  \BibitemOpen
  \bibfield  {author} {\bibinfo {author} {\bibfnamefont {S.}~\bibnamefont
  {Flach}}\ and\ \bibinfo {author} {\bibfnamefont {C.~R.}\ \bibnamefont
  {Willis}},\ }\href@noop {} {\bibfield  {journal} {\bibinfo  {journal} {Phys.
  Rep.}\ }\textbf {\bibinfo {volume} {295}},\ \bibinfo {pages} {181} (\bibinfo
  {year} {1998})}\BibitemShut {NoStop}%
\bibitem [{\citenamefont {Tamga}\ \emph {et~al.}(1995)\citenamefont {Tamga},
  \citenamefont {Remoissenet},\ and\ \citenamefont {Pouget}}]{Tamga95:75}%
  \BibitemOpen
  \bibfield  {author} {\bibinfo {author} {\bibfnamefont {J.~M.}\ \bibnamefont
  {Tamga}}, \bibinfo {author} {\bibfnamefont {M.}~\bibnamefont {Remoissenet}},
  \ and\ \bibinfo {author} {\bibfnamefont {J.}~\bibnamefont {Pouget}},\
  }\href@noop {} {\bibfield  {journal} {\bibinfo  {journal} {Phys. Rev. Lett.}\
  }\textbf {\bibinfo {volume} {75}},\ \bibinfo {pages} {357} (\bibinfo {year}
  {1995})}\BibitemShut {NoStop}%
\bibitem [{\citenamefont {Decker}\ and\ \citenamefont
  {Kasper}(1957)}]{Decker57:10}%
  \BibitemOpen
  \bibfield  {author} {\bibinfo {author} {\bibfnamefont {D.~F.}\ \bibnamefont
  {Decker}}\ and\ \bibinfo {author} {\bibfnamefont {J.~S.}\ \bibnamefont
  {Kasper}},\ }\href@noop {} {\bibfield  {journal} {\bibinfo  {journal} {Acta
  Cryst.}\ }\textbf {\bibinfo {volume} {10}},\ \bibinfo {pages} {332} (\bibinfo
  {year} {1957})}\BibitemShut {NoStop}%
\bibitem [{\citenamefont {Hill}\ \emph {et~al.}(1956)\citenamefont {Hill},
  \citenamefont {Peiser},\ and\ \citenamefont {Rait}}]{Hill56:9}%
  \BibitemOpen
  \bibfield  {author} {\bibinfo {author} {\bibfnamefont {P.~M.}\ \bibnamefont
  {Hill}}, \bibinfo {author} {\bibfnamefont {H.~S.}\ \bibnamefont {Peiser}}, \
  and\ \bibinfo {author} {\bibfnamefont {J.~R.}\ \bibnamefont {Rait}},\
  }\href@noop {} {\bibfield  {journal} {\bibinfo  {journal} {Acta Cryst.}\
  }\textbf {\bibinfo {volume} {9}},\ \bibinfo {pages} {981} (\bibinfo {year}
  {1956})}\BibitemShut {NoStop}%
\bibitem [{\citenamefont {Allain}\ \emph {et~al.}(1966)\citenamefont {Allain},
  \citenamefont {Boucher}, \citenamefont {Imbert},\ and\ \citenamefont
  {Perrin}}]{Allain66:9}%
  \BibitemOpen
  \bibfield  {author} {\bibinfo {author} {\bibfnamefont {Y.}~\bibnamefont
  {Allain}}, \bibinfo {author} {\bibfnamefont {B.}~\bibnamefont {Boucher}},
  \bibinfo {author} {\bibfnamefont {P.}~\bibnamefont {Imbert}}, \ and\ \bibinfo
  {author} {\bibfnamefont {M.}~\bibnamefont {Perrin}},\ }\href@noop {}
  {\bibfield  {journal} {\bibinfo  {journal} {C. R. Acad. Sc. Paris}\ }\textbf
  {\bibinfo {volume} {9}},\ \bibinfo {pages} {263} (\bibinfo {year}
  {1966})}\BibitemShut {NoStop}%
\bibitem [{\citenamefont {Corliss}\ \emph {et~al.}(1967)\citenamefont
  {Corliss}, \citenamefont {Hastings},\ and\ \citenamefont
  {Kunnmann}}]{Corliss67:160}%
  \BibitemOpen
  \bibfield  {author} {\bibinfo {author} {\bibfnamefont {L.~M.}\ \bibnamefont
  {Corliss}}, \bibinfo {author} {\bibfnamefont {J.~M.}\ \bibnamefont
  {Hastings}}, \ and\ \bibinfo {author} {\bibfnamefont {W.}~\bibnamefont
  {Kunnmann}},\ }\href@noop {} {\bibfield  {journal} {\bibinfo  {journal}
  {Phys. Rev.}\ }\textbf {\bibinfo {volume} {160}},\ \bibinfo {pages} {408}
  (\bibinfo {year} {1967})}\BibitemShut {NoStop}%
\bibitem [{\citenamefont {Watanabe}\ \emph {et~al.}(1967)\citenamefont
  {Watanabe}, \citenamefont {Yamauchi}, \citenamefont {Ohashi}, \citenamefont
  {Sugiomoto},\ and\ \citenamefont {Okada}}]{Watanabe67:22}%
  \BibitemOpen
  \bibfield  {author} {\bibinfo {author} {\bibfnamefont {H.}~\bibnamefont
  {Watanabe}}, \bibinfo {author} {\bibfnamefont {H.}~\bibnamefont {Yamauchi}},
  \bibinfo {author} {\bibfnamefont {M.}~\bibnamefont {Ohashi}}, \bibinfo
  {author} {\bibfnamefont {M.}~\bibnamefont {Sugiomoto}}, \ and\ \bibinfo
  {author} {\bibfnamefont {T.}~\bibnamefont {Okada}},\ }\href@noop {}
  {\bibfield  {journal} {\bibinfo  {journal} {J. Phys. Soc. Japan}\ }\textbf
  {\bibinfo {volume} {22}},\ \bibinfo {pages} {939} (\bibinfo {year}
  {1967})}\BibitemShut {NoStop}%
\bibitem [{\citenamefont {Bertaut}\ \emph {et~al.}(1966)\citenamefont
  {Bertaut}, \citenamefont {Chappert}, \citenamefont {Apostolov},\ and\
  \citenamefont {Semenov}}]{Bertaut10:327}%
  \BibitemOpen
  \bibfield  {author} {\bibinfo {author} {\bibfnamefont {E.~F.}\ \bibnamefont
  {Bertaut}}, \bibinfo {author} {\bibfnamefont {J.}~\bibnamefont {Chappert}},
  \bibinfo {author} {\bibfnamefont {A.}~\bibnamefont {Apostolov}}, \ and\
  \bibinfo {author} {\bibfnamefont {V.}~\bibnamefont {Semenov}},\ }\href@noop
  {} {\bibfield  {journal} {\bibinfo  {journal} {Bull. Soc. franc Miner.
  Crist.}\ }\textbf {\bibinfo {volume} {89}},\ \bibinfo {pages} {206} (\bibinfo
  {year} {1966})}\BibitemShut {NoStop}%
\bibitem [{\citenamefont {Shimizu}\ \emph {et~al.}(2003)\citenamefont
  {Shimizu}, \citenamefont {Matsumoto}, \citenamefont {Goto}, \citenamefont
  {Rao}, \citenamefont {Yoshimura},\ and\ \citenamefont
  {Kosuge}}]{Shimizu03:68}%
  \BibitemOpen
  \bibfield  {author} {\bibinfo {author} {\bibfnamefont {T.}~\bibnamefont
  {Shimizu}}, \bibinfo {author} {\bibfnamefont {T.}~\bibnamefont {Matsumoto}},
  \bibinfo {author} {\bibfnamefont {A.}~\bibnamefont {Goto}}, \bibinfo {author}
  {\bibfnamefont {T.~V.~C.}\ \bibnamefont {Rao}}, \bibinfo {author}
  {\bibfnamefont {K.}~\bibnamefont {Yoshimura}}, \ and\ \bibinfo {author}
  {\bibfnamefont {K.}~\bibnamefont {Kosuge}},\ }\href@noop {} {\bibfield
  {journal} {\bibinfo  {journal} {Phys. Rev. B}\ }\textbf {\bibinfo {volume}
  {68}},\ \bibinfo {pages} {224433} (\bibinfo {year} {2003})}\BibitemShut
  {NoStop}%
\bibitem [{\citenamefont {Mizuno}\ \emph {et~al.}(1998)\citenamefont {Mizuno},
  \citenamefont {Tohyama}, \citenamefont {Maekawa}, \citenamefont {Osafune},
  \citenamefont {Motoyama}, \citenamefont {Eisaki},\ and\ \citenamefont
  {Uchida}}]{Mizuno98:57}%
  \BibitemOpen
  \bibfield  {author} {\bibinfo {author} {\bibfnamefont {Y.}~\bibnamefont
  {Mizuno}}, \bibinfo {author} {\bibfnamefont {T.}~\bibnamefont {Tohyama}},
  \bibinfo {author} {\bibfnamefont {S.}~\bibnamefont {Maekawa}}, \bibinfo
  {author} {\bibfnamefont {T.}~\bibnamefont {Osafune}}, \bibinfo {author}
  {\bibfnamefont {N.}~\bibnamefont {Motoyama}}, \bibinfo {author}
  {\bibfnamefont {H.}~\bibnamefont {Eisaki}}, \ and\ \bibinfo {author}
  {\bibfnamefont {S.}~\bibnamefont {Uchida}},\ }\href@noop {} {\bibfield
  {journal} {\bibinfo  {journal} {Phys. Rev. B}\ }\textbf {\bibinfo {volume}
  {57}},\ \bibinfo {pages} {5326} (\bibinfo {year} {1998})}\BibitemShut
  {NoStop}%
\bibitem [{\citenamefont {Obata}\ \emph {et~al.}(2013)\citenamefont {Obata},
  \citenamefont {Obukuro}, \citenamefont {Matsushima}, \citenamefont
  {Nakamura}, \citenamefont {Arai},\ and\ \citenamefont
  {Kobayashi}}]{Obata13:121}%
  \BibitemOpen
  \bibfield  {author} {\bibinfo {author} {\bibfnamefont {K.}~\bibnamefont
  {Obata}}, \bibinfo {author} {\bibfnamefont {Y.}~\bibnamefont {Obukuro}},
  \bibinfo {author} {\bibfnamefont {S.}~\bibnamefont {Matsushima}}, \bibinfo
  {author} {\bibfnamefont {H.}~\bibnamefont {Nakamura}}, \bibinfo {author}
  {\bibfnamefont {M.}~\bibnamefont {Arai}}, \ and\ \bibinfo {author}
  {\bibfnamefont {K.}~\bibnamefont {Kobayashi}},\ }\href@noop {} {\bibfield
  {journal} {\bibinfo  {journal} {J. Ceram. Soc. Jpn.}\ }\textbf {\bibinfo
  {volume} {121}},\ \bibinfo {pages} {766} (\bibinfo {year}
  {2013})}\BibitemShut {NoStop}%
\bibitem [{\citenamefont {Stock}\ \emph {et~al.}(2009)\citenamefont {Stock},
  \citenamefont {Chapon}, \citenamefont {Adamopoulos}, \citenamefont {Lappas},
  \citenamefont {Giot}, \citenamefont {Taylor}, \citenamefont {Green},
  \citenamefont {Brown},\ and\ \citenamefont {Radaelli}}]{Stock09:103}%
  \BibitemOpen
  \bibfield  {author} {\bibinfo {author} {\bibfnamefont {C.}~\bibnamefont
  {Stock}}, \bibinfo {author} {\bibfnamefont {L.~C.}\ \bibnamefont {Chapon}},
  \bibinfo {author} {\bibfnamefont {O.}~\bibnamefont {Adamopoulos}}, \bibinfo
  {author} {\bibfnamefont {A.}~\bibnamefont {Lappas}}, \bibinfo {author}
  {\bibfnamefont {M.}~\bibnamefont {Giot}}, \bibinfo {author} {\bibfnamefont
  {J.~W.}\ \bibnamefont {Taylor}}, \bibinfo {author} {\bibfnamefont {M.~A.}\
  \bibnamefont {Green}}, \bibinfo {author} {\bibfnamefont {C.~M.}\ \bibnamefont
  {Brown}}, \ and\ \bibinfo {author} {\bibfnamefont {P.~G.}\ \bibnamefont
  {Radaelli}},\ }\href@noop {} {\bibfield  {journal} {\bibinfo  {journal}
  {Phys. Rev. Lett.}\ }\textbf {\bibinfo {volume} {103}},\ \bibinfo {pages}
  {077202} (\bibinfo {year} {2009})}\BibitemShut {NoStop}%
\bibitem [{sup()}]{supp}%
  \BibitemOpen
  \href@noop {} {}\bibinfo {note} {Supplementary information giving a list of
  the experimental conditions and also further low temperature inelastic data.
  Information on open access data is also provided.}\BibitemShut {Stop}%
\bibitem [{\citenamefont {Nambu}\ \emph {et~al.}(2015)\citenamefont {Nambu},
  \citenamefont {Gardner}, \citenamefont {MacLaughlin}, \citenamefont {Stock},
  \citenamefont {Endo}, \citenamefont {Jonas}, \citenamefont {Sato},
  \citenamefont {Nakatsuji},\ and\ \citenamefont {Broholm}}]{Nambu15:115}%
  \BibitemOpen
  \bibfield  {author} {\bibinfo {author} {\bibfnamefont {Y.}~\bibnamefont
  {Nambu}}, \bibinfo {author} {\bibfnamefont {J.~S.}\ \bibnamefont {Gardner}},
  \bibinfo {author} {\bibfnamefont {D.~E.}\ \bibnamefont {MacLaughlin}},
  \bibinfo {author} {\bibfnamefont {C.}~\bibnamefont {Stock}}, \bibinfo
  {author} {\bibfnamefont {H.}~\bibnamefont {Endo}}, \bibinfo {author}
  {\bibfnamefont {S.}~\bibnamefont {Jonas}}, \bibinfo {author} {\bibfnamefont
  {T.~J.}\ \bibnamefont {Sato}}, \bibinfo {author} {\bibfnamefont
  {S.}~\bibnamefont {Nakatsuji}}, \ and\ \bibinfo {author} {\bibfnamefont
  {C.}~\bibnamefont {Broholm}},\ }\href@noop {} {\bibfield  {journal} {\bibinfo
   {journal} {Phys. Rev. Lett.}\ }\textbf {\bibinfo {volume} {115}},\ \bibinfo
  {pages} {127202} (\bibinfo {year} {2015})}\BibitemShut {NoStop}%
\bibitem [{\citenamefont {Stock}\ \emph {et~al.}(2006)\citenamefont {Stock},
  \citenamefont {Buyers}, \citenamefont {Yamani}, \citenamefont {Broholm},
  \citenamefont {Chung}, \citenamefont {Tun}, \citenamefont {Liang},
  \citenamefont {Bonn}, \citenamefont {Hardy},\ and\ \citenamefont
  {Birgeneau}}]{Stock06:73}%
  \BibitemOpen
  \bibfield  {author} {\bibinfo {author} {\bibfnamefont {C.}~\bibnamefont
  {Stock}}, \bibinfo {author} {\bibfnamefont {W.~J.~L.}\ \bibnamefont
  {Buyers}}, \bibinfo {author} {\bibfnamefont {Z.}~\bibnamefont {Yamani}},
  \bibinfo {author} {\bibfnamefont {C.~L.}\ \bibnamefont {Broholm}}, \bibinfo
  {author} {\bibfnamefont {J.~H.}\ \bibnamefont {Chung}}, \bibinfo {author}
  {\bibfnamefont {Z.}~\bibnamefont {Tun}}, \bibinfo {author} {\bibfnamefont
  {R.}~\bibnamefont {Liang}}, \bibinfo {author} {\bibfnamefont
  {D.}~\bibnamefont {Bonn}}, \bibinfo {author} {\bibfnamefont {W.~N.}\
  \bibnamefont {Hardy}}, \ and\ \bibinfo {author} {\bibfnamefont {R.~J.}\
  \bibnamefont {Birgeneau}},\ }\href@noop {} {\bibfield  {journal} {\bibinfo
  {journal} {Phys. Rev. B}\ }\textbf {\bibinfo {volume} {73}},\ \bibinfo
  {pages} {100504(R)} (\bibinfo {year} {2006})}\BibitemShut {NoStop}%
\bibitem [{\citenamefont {Stock}\ \emph {et~al.}(2010)\citenamefont {Stock},
  \citenamefont {Jonas}, \citenamefont {Broholm}, \citenamefont {Nakatsuji},
  \citenamefont {Nambu}, \citenamefont {Onuma}, \citenamefont {Maeno},\ and\
  \citenamefont {Chung}}]{Stock10:105}%
  \BibitemOpen
  \bibfield  {author} {\bibinfo {author} {\bibfnamefont {C.}~\bibnamefont
  {Stock}}, \bibinfo {author} {\bibfnamefont {S.}~\bibnamefont {Jonas}},
  \bibinfo {author} {\bibfnamefont {C.}~\bibnamefont {Broholm}}, \bibinfo
  {author} {\bibfnamefont {S.}~\bibnamefont {Nakatsuji}}, \bibinfo {author}
  {\bibfnamefont {Y.}~\bibnamefont {Nambu}}, \bibinfo {author} {\bibfnamefont
  {K.}~\bibnamefont {Onuma}}, \bibinfo {author} {\bibfnamefont
  {Y.}~\bibnamefont {Maeno}}, \ and\ \bibinfo {author} {\bibfnamefont {J.-H.}\
  \bibnamefont {Chung}},\ }\href@noop {} {\bibfield  {journal} {\bibinfo
  {journal} {Phys. Rev. Lett.}\ }\textbf {\bibinfo {volume} {105}},\ \bibinfo
  {pages} {037402} (\bibinfo {year} {2010})}\BibitemShut {NoStop}%
\bibitem [{\citenamefont {Stock}\ \emph {et~al.}(2008)\citenamefont {Stock},
  \citenamefont {Buyers}, \citenamefont {Yamani}, \citenamefont {Tun},
  \citenamefont {Birgeneau}, \citenamefont {Liang}, \citenamefont {Bonn},\ and\
  \citenamefont {Hardy}}]{Stock08:77}%
  \BibitemOpen
  \bibfield  {author} {\bibinfo {author} {\bibfnamefont {C.}~\bibnamefont
  {Stock}}, \bibinfo {author} {\bibfnamefont {W.~J.~L.}\ \bibnamefont
  {Buyers}}, \bibinfo {author} {\bibfnamefont {Z.}~\bibnamefont {Yamani}},
  \bibinfo {author} {\bibfnamefont {Z.}~\bibnamefont {Tun}}, \bibinfo {author}
  {\bibfnamefont {R.~J.}\ \bibnamefont {Birgeneau}}, \bibinfo {author}
  {\bibfnamefont {R.}~\bibnamefont {Liang}}, \bibinfo {author} {\bibfnamefont
  {D.}~\bibnamefont {Bonn}}, \ and\ \bibinfo {author} {\bibfnamefont {W.~N.}\
  \bibnamefont {Hardy}},\ }\href@noop {} {\bibfield  {journal} {\bibinfo
  {journal} {Phys. Rev. B}\ }\textbf {\bibinfo {volume} {77}},\ \bibinfo
  {pages} {104513} (\bibinfo {year} {2008})}\BibitemShut {NoStop}%
\bibitem [{\citenamefont {Murani}\ and\ \citenamefont
  {Heidemann}(1978)}]{Murani78:41}%
  \BibitemOpen
  \bibfield  {author} {\bibinfo {author} {\bibfnamefont {A.~P.}\ \bibnamefont
  {Murani}}\ and\ \bibinfo {author} {\bibfnamefont {A.}~\bibnamefont
  {Heidemann}},\ }\href@noop {} {\bibfield  {journal} {\bibinfo  {journal}
  {Phys. Rev. Lett.}\ }\textbf {\bibinfo {volume} {41}},\ \bibinfo {pages}
  {1402} (\bibinfo {year} {1978})}\BibitemShut {NoStop}%
\bibitem [{\citenamefont {Brown}(2006)}]{Brown:tables}%
  \BibitemOpen
  \bibfield  {author} {\bibinfo {author} {\bibfnamefont {P.~J.}\ \bibnamefont
  {Brown}},\ }\href@noop {} {\emph {\bibinfo {title} {International Tables of
  Crystallography, Vol C}}}\ (\bibinfo  {publisher} {Kluwer, Dordrecht},\
  \bibinfo {year} {2006})\BibitemShut {NoStop}%
\bibitem [{\citenamefont {Tranquada}\ \emph {et~al.}(1996)\citenamefont
  {Tranquada}, \citenamefont {Buttrey},\ and\ \citenamefont
  {Sachan}}]{Tranquada96:54}%
  \BibitemOpen
  \bibfield  {author} {\bibinfo {author} {\bibfnamefont {J.~M.}\ \bibnamefont
  {Tranquada}}, \bibinfo {author} {\bibfnamefont {D.~J.}\ \bibnamefont
  {Buttrey}}, \ and\ \bibinfo {author} {\bibfnamefont {V.}~\bibnamefont
  {Sachan}},\ }\href@noop {} {\bibfield  {journal} {\bibinfo  {journal} {Phys.
  Rev. B}\ }\textbf {\bibinfo {volume} {54}},\ \bibinfo {pages} {12318}
  (\bibinfo {year} {1996})}\BibitemShut {NoStop}%
\bibitem [{\citenamefont {Mezei}(1980)}]{Mezei:nse}%
  \BibitemOpen
  \bibfield  {author} {\bibinfo {author} {\bibfnamefont {F.}~\bibnamefont
  {Mezei}},\ }\href@noop {} {\emph {\bibinfo {title} {Neutron Spin Echo,
  Lecture Notes in Physics, vol 128}}}\ (\bibinfo  {publisher} {Springer,
  Berlin},\ \bibinfo {year} {1980})\BibitemShut {NoStop}%
\bibitem [{\citenamefont {Pickup}\ \emph {et~al.}(2009)\citenamefont {Pickup},
  \citenamefont {Cywinski}, \citenamefont {Pappas}, \citenamefont {Farago},\
  and\ \citenamefont {Fouquet}}]{Pickup09:102}%
  \BibitemOpen
  \bibfield  {author} {\bibinfo {author} {\bibfnamefont {R.~M.}\ \bibnamefont
  {Pickup}}, \bibinfo {author} {\bibfnamefont {R.}~\bibnamefont {Cywinski}},
  \bibinfo {author} {\bibfnamefont {C.}~\bibnamefont {Pappas}}, \bibinfo
  {author} {\bibfnamefont {B.}~\bibnamefont {Farago}}, \ and\ \bibinfo {author}
  {\bibfnamefont {P.}~\bibnamefont {Fouquet}},\ }\href@noop {} {\bibfield
  {journal} {\bibinfo  {journal} {Phys. Rev. Lett.}\ }\textbf {\bibinfo
  {volume} {102}},\ \bibinfo {pages} {097202} (\bibinfo {year}
  {2009})}\BibitemShut {NoStop}%
\bibitem [{\citenamefont {Pappas}\ \emph {et~al.}(2003)\citenamefont {Pappas},
  \citenamefont {Mezei}, \citenamefont {Ehlers}, \citenamefont {Manuel},\ and\
  \citenamefont {Campbell}}]{Pappas03:68}%
  \BibitemOpen
  \bibfield  {author} {\bibinfo {author} {\bibfnamefont {C.}~\bibnamefont
  {Pappas}}, \bibinfo {author} {\bibfnamefont {F.}~\bibnamefont {Mezei}},
  \bibinfo {author} {\bibfnamefont {G.}~\bibnamefont {Ehlers}}, \bibinfo
  {author} {\bibfnamefont {P.}~\bibnamefont {Manuel}}, \ and\ \bibinfo {author}
  {\bibfnamefont {I.~A.}\ \bibnamefont {Campbell}},\ }\href@noop {} {\bibfield
  {journal} {\bibinfo  {journal} {Phys. Rev. B}\ }\textbf {\bibinfo {volume}
  {68}},\ \bibinfo {pages} {054431} (\bibinfo {year} {2003})}\BibitemShut
  {NoStop}%
\bibitem [{\citenamefont {Villain}(1984)}]{Villain84:52}%
  \BibitemOpen
  \bibfield  {author} {\bibinfo {author} {\bibfnamefont {J.}~\bibnamefont
  {Villain}},\ }\href@noop {} {\bibfield  {journal} {\bibinfo  {journal} {Phys.
  Rev. Lett.}\ }\textbf {\bibinfo {volume} {52}},\ \bibinfo {pages} {1543}
  (\bibinfo {year} {1984})}\BibitemShut {NoStop}%
\bibitem [{\citenamefont {Feng}\ \emph {et~al.}(1995)\citenamefont {Feng},
  \citenamefont {Birgeneau},\ and\ \citenamefont {Hill}}]{Feng95:51}%
  \BibitemOpen
  \bibfield  {author} {\bibinfo {author} {\bibfnamefont {Q.}~\bibnamefont
  {Feng}}, \bibinfo {author} {\bibfnamefont {R.~J.}\ \bibnamefont {Birgeneau}},
  \ and\ \bibinfo {author} {\bibfnamefont {J.~P.}\ \bibnamefont {Hill}},\
  }\href@noop {} {\bibfield  {journal} {\bibinfo  {journal} {Phys. Rev. B}\
  }\textbf {\bibinfo {volume} {51}},\ \bibinfo {pages} {15188} (\bibinfo {year}
  {1995})}\BibitemShut {NoStop}%
\bibitem [{\citenamefont {Nattermann}\ and\ \citenamefont
  {Vilfan}(1988)}]{Natt88:61}%
  \BibitemOpen
  \bibfield  {author} {\bibinfo {author} {\bibfnamefont {T.}~\bibnamefont
  {Nattermann}}\ and\ \bibinfo {author} {\bibfnamefont {I.}~\bibnamefont
  {Vilfan}},\ }\href@noop {} {\bibfield  {journal} {\bibinfo  {journal} {Phys.
  Rev. Lett.}\ }\textbf {\bibinfo {volume} {61}},\ \bibinfo {pages} {223}
  (\bibinfo {year} {1988})}\BibitemShut {NoStop}%
\bibitem [{\citenamefont {Coldea}\ \emph {et~al.}(2010)\citenamefont {Coldea},
  \citenamefont {Tennant}, \citenamefont {Wheeler}, \citenamefont {Wawrzynska},
  \citenamefont {Prabhakaran}, \citenamefont {Telling}, \citenamefont
  {Habicht}, \citenamefont {Smeibidl},\ and\ \citenamefont
  {Kiefer}}]{Coldea10:327}%
  \BibitemOpen
  \bibfield  {author} {\bibinfo {author} {\bibfnamefont {R.}~\bibnamefont
  {Coldea}}, \bibinfo {author} {\bibfnamefont {D.~A.}\ \bibnamefont {Tennant}},
  \bibinfo {author} {\bibfnamefont {E.~M.}\ \bibnamefont {Wheeler}}, \bibinfo
  {author} {\bibfnamefont {E.}~\bibnamefont {Wawrzynska}}, \bibinfo {author}
  {\bibfnamefont {D.}~\bibnamefont {Prabhakaran}}, \bibinfo {author}
  {\bibfnamefont {M.}~\bibnamefont {Telling}}, \bibinfo {author} {\bibfnamefont
  {K.}~\bibnamefont {Habicht}}, \bibinfo {author} {\bibfnamefont
  {P.}~\bibnamefont {Smeibidl}}, \ and\ \bibinfo {author} {\bibfnamefont
  {K.}~\bibnamefont {Kiefer}},\ }\href@noop {} {\bibfield  {journal} {\bibinfo
  {journal} {Science}\ }\textbf {\bibinfo {volume} {327}},\ \bibinfo {pages}
  {177} (\bibinfo {year} {2010})}\BibitemShut {NoStop}%
\bibitem [{\citenamefont {Morris}\ \emph {et~al.}(2014)\citenamefont {Morris},
  \citenamefont {Aguilar}, \citenamefont {Ghosh}, \citenamefont {Koohpayeh},
  \citenamefont {Krizan}, \citenamefont {Cava}, \citenamefont {Tchernyshyov},
  \citenamefont {McWueen},\ and\ \citenamefont {Armitage}}]{Morris14:112}%
  \BibitemOpen
  \bibfield  {author} {\bibinfo {author} {\bibfnamefont {C.~M.}\ \bibnamefont
  {Morris}}, \bibinfo {author} {\bibfnamefont {R.~V.}\ \bibnamefont {Aguilar}},
  \bibinfo {author} {\bibfnamefont {A.}~\bibnamefont {Ghosh}}, \bibinfo
  {author} {\bibfnamefont {S.~M.}\ \bibnamefont {Koohpayeh}}, \bibinfo {author}
  {\bibfnamefont {J.}~\bibnamefont {Krizan}}, \bibinfo {author} {\bibfnamefont
  {R.~J.}\ \bibnamefont {Cava}}, \bibinfo {author} {\bibfnamefont
  {O.}~\bibnamefont {Tchernyshyov}}, \bibinfo {author} {\bibfnamefont {T.~M.}\
  \bibnamefont {McWueen}}, \ and\ \bibinfo {author} {\bibfnamefont {N.~P.}\
  \bibnamefont {Armitage}},\ }\href@noop {} {\bibfield  {journal} {\bibinfo
  {journal} {Phys. Rev. Lett.}\ }\textbf {\bibinfo {volume} {112}},\ \bibinfo
  {pages} {137403} (\bibinfo {year} {2014})}\BibitemShut {NoStop}%
\bibitem [{\citenamefont {Judd}(1998)}]{Judd:book}%
  \BibitemOpen
  \bibfield  {author} {\bibinfo {author} {\bibfnamefont {B.~R.}\ \bibnamefont
  {Judd}},\ }\href@noop {} {\emph {\bibinfo {title} {Operator Techniques in
  Atomic Spectroscopy}}}\ (\bibinfo  {publisher} {Princeton University Press,
  New Jersey},\ \bibinfo {year} {1998})\BibitemShut {NoStop}%
\bibitem [{\citenamefont {Sato}\ and\ \citenamefont
  {Sievers}(2004)}]{Sato04:432}%
  \BibitemOpen
  \bibfield  {author} {\bibinfo {author} {\bibfnamefont {M.}~\bibnamefont
  {Sato}}\ and\ \bibinfo {author} {\bibfnamefont {A.~J.}\ \bibnamefont
  {Sievers}},\ }\href@noop {} {\bibfield  {journal} {\bibinfo  {journal}
  {Nature}\ }\textbf {\bibinfo {volume} {432}},\ \bibinfo {pages} {486}
  (\bibinfo {year} {2004})}\BibitemShut {NoStop}%
\bibitem [{\citenamefont {Sakaguchi}\ and\ \citenamefont
  {Malomed}(2005)}]{Sakaguchi05:72}%
  \BibitemOpen
  \bibfield  {author} {\bibinfo {author} {\bibfnamefont {H.}~\bibnamefont
  {Sakaguchi}}\ and\ \bibinfo {author} {\bibfnamefont {B.~A.}\ \bibnamefont
  {Malomed}},\ }\href@noop {} {\bibfield  {journal} {\bibinfo  {journal} {Phys.
  Rev. E}\ }\textbf {\bibinfo {volume} {72}},\ \bibinfo {pages} {046610}
  (\bibinfo {year} {2005})}\BibitemShut {NoStop}%
\bibitem [{\citenamefont {Flach}\ and\ \citenamefont
  {Gorbach}(2008)}]{Flach08:497}%
  \BibitemOpen
  \bibfield  {author} {\bibinfo {author} {\bibfnamefont {S.}~\bibnamefont
  {Flach}}\ and\ \bibinfo {author} {\bibfnamefont {A.~V.}\ \bibnamefont
  {Gorbach}},\ }\href@noop {} {\bibfield  {journal} {\bibinfo  {journal} {Phys.
  Rep.}\ }\textbf {\bibinfo {volume} {467}},\ \bibinfo {pages} {1} (\bibinfo
  {year} {2008})}\BibitemShut {NoStop}%
\bibitem [{\citenamefont {Mikeska}(1978)}]{Mikeska78:11}%
  \BibitemOpen
  \bibfield  {author} {\bibinfo {author} {\bibfnamefont {M.~J.}\ \bibnamefont
  {Mikeska}},\ }\href@noop {} {\bibfield  {journal} {\bibinfo  {journal} {J.
  Phys. C: Solid State Phys.}\ }\textbf {\bibinfo {volume} {11}},\ \bibinfo
  {pages} {L29} (\bibinfo {year} {1978})}\BibitemShut {NoStop}%
\bibitem [{\citenamefont {Haldane}(1983)}]{Haldane83:50}%
  \BibitemOpen
  \bibfield  {author} {\bibinfo {author} {\bibfnamefont {F.~D.~M.}\
  \bibnamefont {Haldane}},\ }\href@noop {} {\bibfield  {journal} {\bibinfo
  {journal} {Phys. Rev. Lett.}\ }\textbf {\bibinfo {volume} {50}},\ \bibinfo
  {pages} {1153} (\bibinfo {year} {1983})}\BibitemShut {NoStop}%
\bibitem [{\citenamefont {Fogedby}(1980)}]{Fogedby80:13}%
  \BibitemOpen
  \bibfield  {author} {\bibinfo {author} {\bibfnamefont {H.~C.}\ \bibnamefont
  {Fogedby}},\ }\href@noop {} {\bibfield  {journal} {\bibinfo  {journal} {J.
  Phys. A: Math. Gen.}\ }\textbf {\bibinfo {volume} {13}},\ \bibinfo {pages}
  {1467} (\bibinfo {year} {1980})}\BibitemShut {NoStop}%
\bibitem [{\citenamefont {Haldane}(1982)}]{Haldane82:15}%
  \BibitemOpen
  \bibfield  {author} {\bibinfo {author} {\bibfnamefont {F.~D.~M.}\
  \bibnamefont {Haldane}},\ }\href@noop {} {\bibfield  {journal} {\bibinfo
  {journal} {J. Phys. C: Solid State Phys.}\ }\textbf {\bibinfo {volume}
  {15}},\ \bibinfo {pages} {L1309} (\bibinfo {year} {1982})}\BibitemShut
  {NoStop}%
\bibitem [{\citenamefont {Pushkarov}\ and\ \citenamefont
  {Pushkarov}(1977)}]{Pushkarov77:81}%
  \BibitemOpen
  \bibfield  {author} {\bibinfo {author} {\bibfnamefont {D.~I.}\ \bibnamefont
  {Pushkarov}}\ and\ \bibinfo {author} {\bibfnamefont {K.~I.}\ \bibnamefont
  {Pushkarov}},\ }\href@noop {} {\bibfield  {journal} {\bibinfo  {journal}
  {Phys. Stat. Sol. (b)}\ }\textbf {\bibinfo {volume} {81}},\ \bibinfo {pages}
  {703} (\bibinfo {year} {1977})}\BibitemShut {NoStop}%
\bibitem [{\citenamefont {Pushkarov}\ and\ \citenamefont
  {Pushkarov}(1978)}]{Pushkarov78:85}%
  \BibitemOpen
  \bibfield  {author} {\bibinfo {author} {\bibfnamefont {D.~I.}\ \bibnamefont
  {Pushkarov}}\ and\ \bibinfo {author} {\bibfnamefont {K.~I.}\ \bibnamefont
  {Pushkarov}},\ }\href@noop {} {\bibfield  {journal} {\bibinfo  {journal}
  {Phys. Stat. Sol. (b)}\ }\textbf {\bibinfo {volume} {85}},\ \bibinfo {pages}
  {K89} (\bibinfo {year} {1978})}\BibitemShut {NoStop}%
\bibitem [{\citenamefont {Pushkarov}\ and\ \citenamefont
  {Pushkarov}(1979)}]{Pushkarov79:93}%
  \BibitemOpen
  \bibfield  {author} {\bibinfo {author} {\bibfnamefont {D.~I.}\ \bibnamefont
  {Pushkarov}}\ and\ \bibinfo {author} {\bibfnamefont {K.~I.}\ \bibnamefont
  {Pushkarov}},\ }\href@noop {} {\bibfield  {journal} {\bibinfo  {journal}
  {Phys. Stat. Sol. (b)}\ }\textbf {\bibinfo {volume} {93}},\ \bibinfo {pages}
  {735} (\bibinfo {year} {1979})}\BibitemShut {NoStop}%
\bibitem [{\citenamefont {Mikeska}\ and\ \citenamefont
  {Steiner}(1991)}]{Mikeska91:40}%
  \BibitemOpen
  \bibfield  {author} {\bibinfo {author} {\bibfnamefont {M.~J.}\ \bibnamefont
  {Mikeska}}\ and\ \bibinfo {author} {\bibfnamefont {M.}~\bibnamefont
  {Steiner}},\ }\href@noop {} {\bibfield  {journal} {\bibinfo  {journal} {Adv.
  Phys.}\ }\textbf {\bibinfo {volume} {40}},\ \bibinfo {pages} {191} (\bibinfo
  {year} {1991})}\BibitemShut {NoStop}%
\bibitem [{\citenamefont {Marquie}\ \emph {et~al.}(1995)\citenamefont
  {Marquie}, \citenamefont {Bilbault},\ and\ \citenamefont
  {Remoissenet}}]{Marquie95:51}%
  \BibitemOpen
  \bibfield  {author} {\bibinfo {author} {\bibfnamefont {P.}~\bibnamefont
  {Marquie}}, \bibinfo {author} {\bibfnamefont {J.~M.}\ \bibnamefont
  {Bilbault}}, \ and\ \bibinfo {author} {\bibfnamefont {M.}~\bibnamefont
  {Remoissenet}},\ }\href@noop {} {\bibfield  {journal} {\bibinfo  {journal}
  {Phys. Rev. E}\ }\textbf {\bibinfo {volume} {51}},\ \bibinfo {pages} {6127}
  (\bibinfo {year} {1995})}\BibitemShut {NoStop}%
\bibitem [{\citenamefont {Denardo}\ \emph {et~al.}(1992)\citenamefont
  {Denardo}, \citenamefont {Galvin}, \citenamefont {Greenfield}, \citenamefont
  {Larraza}, \citenamefont {Putterman},\ and\ \citenamefont
  {Wright}}]{Denardo92:68}%
  \BibitemOpen
  \bibfield  {author} {\bibinfo {author} {\bibfnamefont {B.}~\bibnamefont
  {Denardo}}, \bibinfo {author} {\bibfnamefont {B.}~\bibnamefont {Galvin}},
  \bibinfo {author} {\bibfnamefont {A.}~\bibnamefont {Greenfield}}, \bibinfo
  {author} {\bibfnamefont {A.}~\bibnamefont {Larraza}}, \bibinfo {author}
  {\bibfnamefont {S.}~\bibnamefont {Putterman}}, \ and\ \bibinfo {author}
  {\bibfnamefont {W.}~\bibnamefont {Wright}},\ }\href@noop {} {\bibfield
  {journal} {\bibinfo  {journal} {Phys. Rev. Lett.}\ }\textbf {\bibinfo
  {volume} {68}},\ \bibinfo {pages} {1730} (\bibinfo {year}
  {1992})}\BibitemShut {NoStop}%
\bibitem [{\citenamefont {Eisenberg}\ \emph {et~al.}(1998)\citenamefont
  {Eisenberg}, \citenamefont {Silberg}, \citenamefont {Morandotti},
  \citenamefont {Boyd},\ and\ \citenamefont {Aitchison}}]{Eisenberg98:81}%
  \BibitemOpen
  \bibfield  {author} {\bibinfo {author} {\bibfnamefont {H.~S.}\ \bibnamefont
  {Eisenberg}}, \bibinfo {author} {\bibfnamefont {Y.}~\bibnamefont {Silberg}},
  \bibinfo {author} {\bibfnamefont {R.}~\bibnamefont {Morandotti}}, \bibinfo
  {author} {\bibfnamefont {A.~R.}\ \bibnamefont {Boyd}}, \ and\ \bibinfo
  {author} {\bibfnamefont {J.~S.}\ \bibnamefont {Aitchison}},\ }\href@noop {}
  {\bibfield  {journal} {\bibinfo  {journal} {Phys. Rev. Lett.}\ }\textbf
  {\bibinfo {volume} {81}},\ \bibinfo {pages} {3383} (\bibinfo {year}
  {1998})}\BibitemShut {NoStop}%
\bibitem [{\citenamefont {Sato}\ and\ \citenamefont
  {Sievers}(2005)}]{Sato05:71}%
  \BibitemOpen
  \bibfield  {author} {\bibinfo {author} {\bibfnamefont {M.}~\bibnamefont
  {Sato}}\ and\ \bibinfo {author} {\bibfnamefont {A.~J.}\ \bibnamefont
  {Sievers}},\ }\href@noop {} {\bibfield  {journal} {\bibinfo  {journal} {Phys.
  Rev. B}\ }\textbf {\bibinfo {volume} {71}},\ \bibinfo {pages} {214306}
  (\bibinfo {year} {2005})}\BibitemShut {NoStop}%
\end{thebibliography}
%\end{thebibliography}

%merlin.mbs apsrev4-1.bst 2010-07-25 4.21a (PWD, AO, DPC) hacked
%Control: key (0)
%Control: author (8) initials jnrlst
%Control: editor formatted (1) identically to author
%Control: production of article title (-1) disabled
%Control: page (0) single
%Control: year (1) truncated
%Control: production of eprint (0) enabled
%

\end{document}

\title{Supplementary information for ``Solitary magnons in the $S={5\over2}$ antiferromagnet-CaFe$_{2}$O$_{4}$"}

\author{C. Stock}
\affiliation{School of Physics and Astronomy and Centre for Science at Extreme Conditions, University of Edinburgh, Edinburgh EH9 3FD, UK}

\author{E. E. Rodriguez}
\affiliation{Department of Chemistry and Biochemistry, University of Maryland, College Park, Maryland 20742, USA}

\author{N. Lee}
\affiliation{Rutgers Center for Emergent Materials and Department of Physics and Astronomy, Rutgers University, 136 Frelinghuysen Road, Piscataway, New Jersey 08854, USA}

\author{M. A. Green}
\affiliation{School of Physical Sciences, University of Kent, Canterbury, CT2 7NH, UK}

\author{F. Demmel}
\affiliation{ISIS Facility, Rutherford Appleton Labs, Chilton, Didcot, OX11 0QX, UK}

\author{R. A. Ewings}
\affiliation{ISIS Facility, Rutherford Appleton Labs, Chilton, Didcot, OX11 0QX, UK}

\author{P. Fouquet}
\affiliation{Institute Laue-Langevin, 6 rue Jules Horowitz, Boite Postale 156, 38042 Grenoble Cedex 9, France}

\author{M. Laver}
\affiliation{Laboratory for Neutron Scattering, Paul Scherrer Institut, CH-5232 Villigen, Switzerland}

\author{Ch. Niedermayer}
\affiliation{Laboratory for Neutron Scattering, Paul Scherrer Institut, CH-5232 Villigen, Switzerland}

\author{Y. Su}
\author{K. Nemkovski}
\affiliation{J\"ulich Centre for Neuton Science JCNS, Forschungszentrum J\"ulich GmbH, Outstation at MLZ, Lichtenbergstra\ss e 1, D-85747 Garching, Germany}

\author{J. A. Rodriguez-Rivera}
\affiliation{NIST Center for Neutron Research, National Institute of Standards and Technology, 100 Bureau Drive, Gaithersburg, Maryland, 20899, USA}
\affiliation{Department of Materials Science, University of Maryland, College Park, Maryland 20742, USA}

\author{S. -W. Cheong}
\affiliation{Rutgers Center for Emergent Materials and Department of Physics and Astronomy, Rutgers University, 136 Frelinghuysen Road, Piscataway, New Jersey 08854, USA}

\date{\today}

\begin{abstract}

Supplementary information is provided regarding the experimental configuration and open access data.  

\end{abstract}

\pacs{}

\maketitle

%\section{Experimental information:}  

%\subsection{Neutron Instruments:}

To understand the origin of the confined spin excitations, we have applied a number of different neutron diffraction and spectroscopy techniques.  In this section we outline the details of the experimental configurations for these measurements.

\textit{SPINS and  BT9 (NIST):} Single crystal diffraction data measuring the magnetic structure as a function of temperature were performed on the SPINS and BT9 triple-axis spectrometers (NIST, USA).  Vertically focussed PG002 graphite was used to produce a monochromatic beam incident on the sample and a flat PG002 graphite analyzer was used on the scattered side to select a final energy and reduce background.  For SPINS the final energy was fixed at E$_{f}$=5.0 meV and BT9 E$_{f}$=13.7 meV.  Cold beryllium filters were used to remove higher order contamination on SPINS.  A graphite filter was used on BT9 on both the incident and scattered sides.  For experiments done on both instruments, the collimation sequence was set to $open-80-S-80-open$.

\begin{figure}[t]
\includegraphics[width=9.2cm]{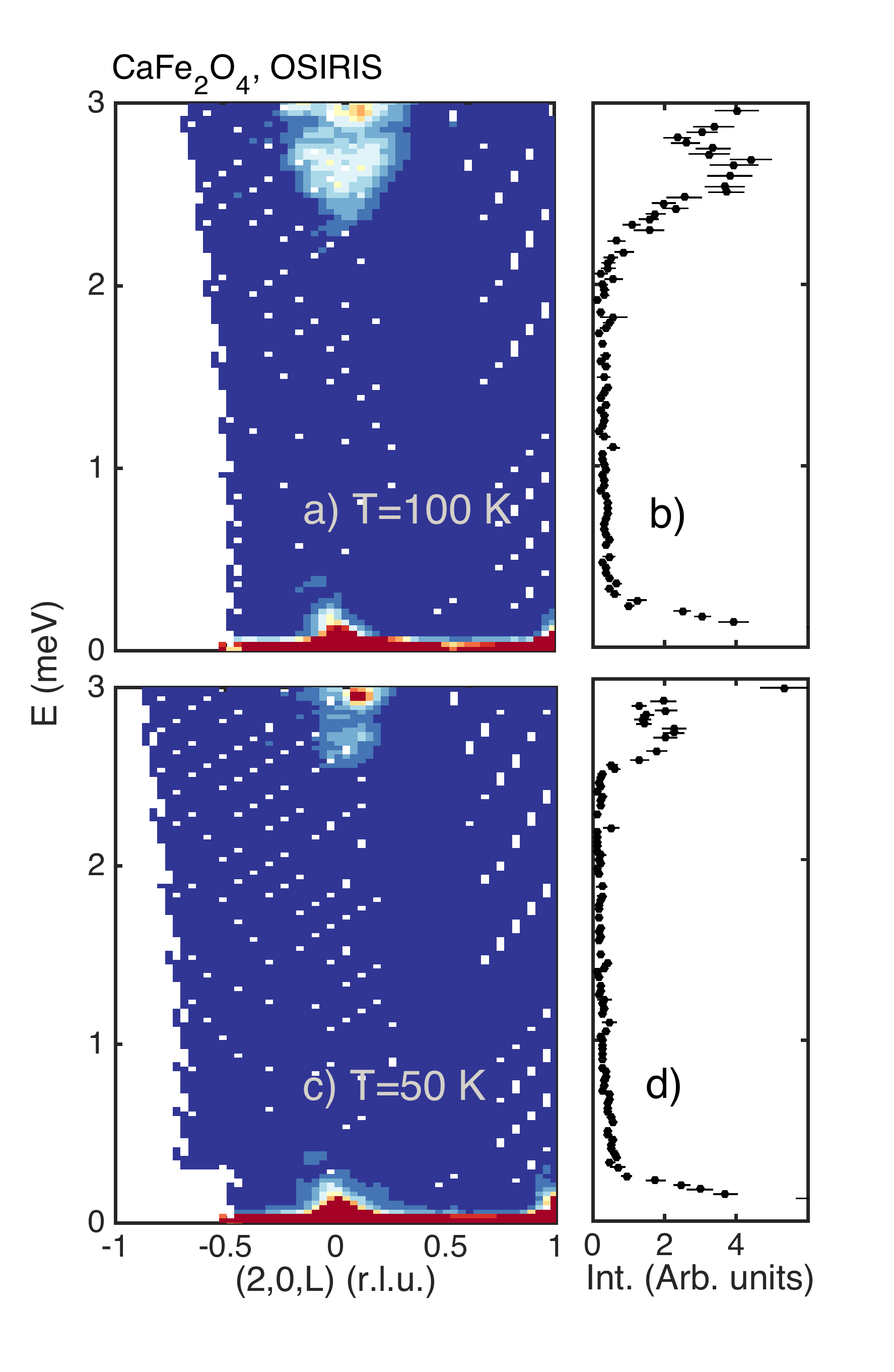}
\caption{\label{suppl} Constant-Q slices at $(a)$ 100 K and $(c)$ 50 K.  Panels $(b)$ and $(d)$ show momentum integrated energy scans with L=[-0.1,0.1] r.l.u.  The quantized excitations are present to the lowest temperatures measured on OSIRIS.}
\end{figure}

\textit{DCS (NIST):} To map out the low-energy fluctuations and extract the $c$-axis dispersion, we used the DCS chopper spectrometer with E$_{i}$=14.1 meV.  Measurements were made in the (H0L) scattering plane with each orientation spaced 1$^{\circ}$ apart.  The DCS instrument consists of 325 detectors in the horizontal scattering plane and this configuration afforded an energy resolution (full-width) of $2\delta E$=1.2 meV.  For the data presented in the main text, we note that the out of plane detector banks were masked to consider scattering truely within the (H0L) plane.

\textit{RITA-2 (PSI):} Complementing our data from SPINS and BT9, we have also presented neutron elastic and inelastic data from RITA-2.  For these measurements, we have used the position sensitive detector which is well suited for tracking diffuse scattering and studying dispersive excitations.   Similar to SPINS, a vertically focussed PG002 monochromator was used along with a flat PG002 analyzer.  The final energy was fixed to E$_{f}$=4.5 meV and dynamics were studied by varying E$_{i}$ therefore defining the energy transfer as $E=E_{i}-E_{f}$.  The background from higher order scattering was reduced using a cold Beryllium filter on the scattered side.  This configuration gave an energy resolution (full-width) of $2\delta E$=0.21 meV at the elastic position.  Calibrations for the absolute moments of the $A$ and $B$ phases were done using RITA-2 by comparing the (1,0,1) and (1,0,2) Bragg peaks against 12 nuclear Bragg peaks where magnetic scattering was absent.  Crystallographic data for the nuclear structure factors was confirmed with powder experiments on the BT1 powder diffractometer at NIST.

\textit{DNS (FRM2)}: To confirm the magnetic structure and its relation to the magnetic structure, we used the DNS diffractometer at FRM2 (Germany).  Measurements were made with E$_{i}$=4.6 meV in two-axis mode (energy integrating) therefore providing an approximate measure $S(\vec{Q})$.  The incident beam is selected using a horizontally and vertically focused PG002 monochromator.  The beam was then polarized used m=3 Scharpf supermirror polarizers.  The polarization at the sample was fixed through the use of an $xyz$ coil with the $x$ direction chosen to be parallel to the average $\vec{Q}$ at the sample and the $z$ vertical.  With the use of flipping coils in the incident and scattered beams, the two spin-flip and non-spin-flip cross sections could be measured with the neutron polarization along the three orthogonal Cartesian coordinates.  The flipping ratio was 20 $\pm$ 1 and was not found to deviate from this value regardless of the direction of neutron polarization.  All spin-flip data have been corrected for the feed-through from the non-spin-lip channel.  The scattered beam was measured with 24 detectors equally spaced $5^{\circ}$ aprt covering a total angular range in scattering angle equal to 120$^{\circ}$.   

 \textit{MAPS (ISIS):} To obtain the dispersion in the (HK0) scattering plane where the dispersion is the strongest, we have used the MAPS time of flight spectrometer at ISIS (UK).  Unlike DCS, MAPS has position sensitive detectors allowing measurements out of the scattering plane to be obtained.  Taking advantage of the weak $c$-axis dispersion found on DCS, the CaFe$_{2}$O$_{4}$ crystal was aligned such that the $c$-axis was oriented along the incident beam.  The incident energy was fixed E$_{i}$=75 meV using a Fermi chopper spun at 200 Hz.  Background from high energy neutrons was reduced using a $t_{0}$ chopper spun at 50 Hz.  This configuration gave an elastic energy resolution (full-width) of  $2\delta E$=4.0 meV.
 
 \textit{OSIRIS (ISIS):} High momentum and energy resolution data was obtained using the OSIRIS backscattering spectrometer located at ISIS.  A whitebeam of neutrons is incident on the sample and the final energy is fixed at E$_{f}$=1.84 meV using cooled graphite analyzers.  A cooled Beryllium filter was used on the scattered side to reduce background.  The default time configuration is set for a dynamic range of $\pm$ 0.5 meV, however by delaying the time channels, the dynamic range was extended to 3 meV.   An elastic energy resolution (full-width) of $2\delta E$=0.025 meV was obtained for these experiments.  Due to kinematic constraints, we focussed our measurements around $\vec{Q}\sim (2,0,0)$ so that the quantized excitations could be tracked up to energy transfers of $\sim$ 3 meV.   As illustrated in Fig. \ref{suppl}, quantized excitations are observed to low temperatures as the overall envelope of the spin-waves recovers to higher energy transfers.

\textit{IN11 (ILL):} Spin-echo measurements were done on the IN11 spectrometer (ILL, France) probing the low-energy $\sim$ GHz spin response.  Neutron spin-echo spectroscopy differs from other neutron methods in that it measures the real part of the normalized intermediate scattering.   This is achieved by encoding the neutron's speed into the Larmor precession of its nuclear magnetic moment in a well controlled, externally applied magnetic field.  $I(Q,t)$ is the spatial Fourier transform of the Van Hove self correlation function $G(r,t)$ which, essentially, gives the probability of finding a particle after time $t$ at a radius $r$ around the original position.   To obtain the dynamic range in time shown in the main text, two wavelengths were used - 5.5 \AA\ and 4.2 \AA.  

\textit{Open data access:} Following UK research council guidance, data files can be accessed either at source (from the ILL (www.ill.eu), or the NCNR (www.ncnr.nist.gov) or through the University of Edinburgh's online digital repository (datashare.is.ed.ac.uk) after publication.

\bibliography{CFO_bib}